\documentclass[10pt]{article}
\usepackage[left=1.8cm,right=1.8cm,top=2cm,bottom=1.8cm]{geometry}
\usepackage[english]{babel}
\usepackage[utf8]{inputenc}

\usepackage[pdftex]{graphicx}
\DeclareGraphicsExtensions{.pdf}
\usepackage{amsmath}
\usepackage{hyperref}

\usepackage{tikz}
\usetikzlibrary{arrows.meta}

\providecommand\Rey{\textit{Re}}
\let\ig\includegraphics
\let\tw\textwidth
\let\lw\linewidth

\author{Gonzalo Arranz$^{a,*}$, 
    Yuenong Ling$^{a}$, Sam Costa$^{a}$, Konrad Goc$^{b}$ 
    \& Adri\'an Lozano-Dur\'{a}n$^{a,*}$\\
    \small $^{a}$Department of Aeronautics and Astronautics, Massachusetts
    Institute of Technology, Cambridge, MA 02139\\
    \small $^{a}$The Boeing Company, Everett, 98204, WA, USA\\
    \small $^{*}$garranz@mit.edu or adrianld@mit.edu
    }

\date{}

\title{Building-block flow model for large-eddy simulation}

\begin{document}

\maketitle
\begin{abstract}
We introduce a closure model for wall-modeled large-eddy simulation (WMLES),
referred to as the Building-block Flow Model (BFM). The foundation of
the model rests on the premise that a finite collection of simple
flows encapsulates the essential physics necessary to predict more
complex scenarios. The BFM is implemented using artificial neural
networks and introduces five  advancements within the framework
of WMLES: (1) It is designed to predict multiple flow
regimes (wall turbulence under zero, favorable, adverse
mean-pressure-gradient, and separation); (2) It unifies the closure
model at solid boundaries (i.e., the wall model) and the rest of the
flow (i.e., the subgrid-scale model) into a single entity; (3) It
ensures consistency with numerical schemes and gridding strategy by
accounting for numerical errors; (4) It is directly applicable to
arbitrary complex geometries; (5) It can be scaled up to model
additional flow physics in the future if needed (e.g., shockwaves and
laminar-to-turbulent transition). The BFM is utilized to predict key
quantities of interest in turbulent channel and pipe flows, a Gaussian bump,
a simplified aircraft, and a realistic aircraft in landing configuration. In all cases, the BFM
demonstrates similar or superior capabilities in terms of accuracy and
computational efficiency compared to previous state-of-the-art closure
models.
\end{abstract}

\section{Introduction}
The adoption of transformative low-emissions aero/hydro vehicle
designs and propulsion systems is significantly impeded by a major
obstacle: the lengthy and expensive experimental campaigns needed
throughout the design cycle. These campaigns can span years and cost
billions of dollars~\cite{mauery2021}. Virtual testing via
computational fluid dynamics (CFD) might accelerate the process and
alleviate costs~\cite{mani2023}. However, current CFD closure models
do not comply with the stringent accuracy requirements demanded by the
aerospace industry~\cite{slotnick2014}. These limitations primarily
stem from the challenging nature of turbulence, i.e., the chaotic and
multiscale motion of flows, resulting in complex physical phenomena.

Despite the challenges, novel modeling strategies hold the potential
to unlock substantial financial benefits worldwide, extending far
beyond the aerospace industry.  The use of highly accurate CFD tools
can facilitate the creation of innovative vehicle designs that achieve
up to 30\% fuel savings~\cite{dangelo2010n}. This breakthrough would
result in substantial cost reductions, amounting to hundreds of
billions of dollars per year within the transportation sector alone.
For instance, the global ocean shipping industry consumes
approximately 2 billion barrels of oil annually~\cite{smits2013}.
Similarly, the airline and trucking industries consume approximately
1.5 billion barrels and 1.2 billion barrels of oil per year,
respectively.  Additionally, a mere 5\% reduction in transportation
drag is estimated to have an impact equivalent to doubling the current
wind energy production in the United States~\cite{smits2013}. The
implementation of novel turbulence modeling techniques also carries
immense potential for mitigating environmental issues and reducing
pollutant emissions~\cite{smith2013us}.

\begin{figure}[t]
  \centering
  \definecolor{myC0}{HTML}{003F5C}
  \definecolor{myC1}{HTML}{D92f2f}
  \definecolor{myC2}{HTML}{FFA600}
  
  \begin{tikzpicture}[font=\sffamily]
      \node[anchor=south west] (f1a) at (0,0) {\ig[width=.5\tw]{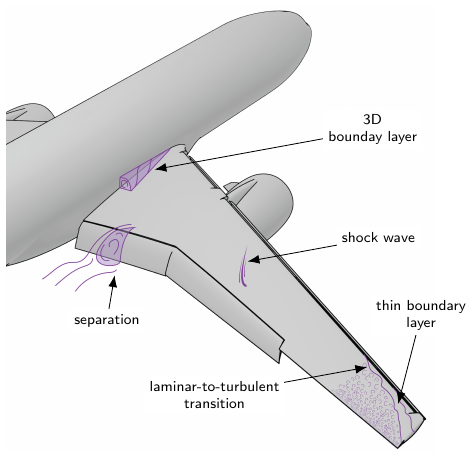}};
      \node[anchor=north west] at (f1a.north east) 
      {\ig[width=.49\tw]{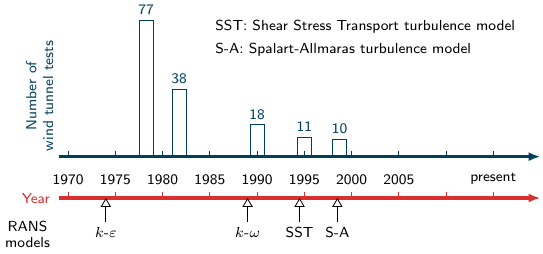}};
      
      \pgfmathsetmacro{\hlen}{1.1}
      \pgfmathsetmacro{\vtex}{.3}
      \begin{scriptsize}
          \begin{scope}[xshift=.5\tw, yshift=+12em,scale=1.3]
              \node[anchor=south west] at (.8,.5) {\textbf{Selected milestones in WMLES}};
          \draw[ultra thick] (.8,0) -- (7,0);
          \draw[ultra thick,->] (.8,-\hlen) -- (7.,-\hlen);
  
          \foreach \i in {1,2,...,6}{ 
              \draw[myC0] (1*\i,   0) --+ (0,+.1);
              \draw[myC0] (1*\i+.5, -\hlen) --+ (0,+.1);
          }
  
          \foreach[count=\i] \y in {1965,1970,...,1990}
              { \node[above,yshift=.5em] at (\i,0) {\y}; }
          \foreach[count=\i] \y in {1995,2000,...,2020}
              { \node[above,yshift=.5em] at (\i+.5,-\hlen) {\y}; }
  
          \draw[black,{Triangle[fill=white,scale=1.2]}-] (.9,0) --++ (0,-\vtex) 
              node [below,fill=myC0!10] {Smag.\cite{smagorinsky1963}};
          \draw[black,{Triangle[fill=white,scale=1.2]}-] (2,0) --++ (0,-\vtex) 
              node [below,fill=myC0!10] {LoW-WM\cite{deardorff1970}};
          \draw[black,{Triangle[fill=white,scale=1.2]}-] (4,0) --++ (0,-\vtex) 
              node [below,fill=myC0!10] {SSM\cite{bardina1980}};
          \draw[black,{Triangle[fill=white,scale=1.2]}-] (6.2,0) --++ (0,-\vtex) 
              node [below,fill=myC0!10] {DSM\cite{germano1991}};
          \draw[black,{Triangle[fill=white,scale=1.2]}-] (.5+1.8,-\hlen) --++ (0,-\vtex) 
              node [anchor=north east,fill=myC0!10,xshift=5pt] {WALE\cite{nicoud1999},
              Deconv.-SGS\cite{stolz1999}};
          \draw[black,{Triangle[fill=white,scale=1.2]}-] (.5+2.0,-\hlen) --++
              (0,-2.2*\vtex) 
              node [below,fill=myC0!10,anchor=north east,xshift=+5pt]
              {TBL-WM\cite{cabot2000}};
          \draw[black,{Triangle[fill=white,scale=1.2]}-] (.5+2.4,-\hlen) --++
              (0,-3.3*\vtex) 
              node [below,fill=myC1!10] {ANN SGS\cite{sarghini2003}};
          \draw[black,{Triangle[fill=white,scale=1.2]}-] (.5+2.9,-\hlen) --++ (0,-\vtex) 
              node [below,fill=myC0!10] {Vreman\cite{vreman2004}};
          \draw[black,{Triangle[fill=white,scale=1.2]}-] (.5+4.4,-\hlen) --++ (0,-\vtex) 
              node [below,fill=myC0!10] {EQ-WM\cite{kawai2012}};
          \draw[black,{Triangle[fill=white,scale=1.2]}-] (.5+5.8,-\hlen) --++ (0,-\vtex) 
              node [below,fill=myC1!10] {ANN WM\cite{yang2019predictive}};
          \draw[black,{Triangle[fill=white,scale=1.2]}-] (.5+5.2,-\hlen) --++
              (0,-2.2*\vtex) 
              node [below,fill=myC0!10,text width=4.5em,align=center] 
              {WMLES of an \\ aircraft\cite{lehmkuhl2016}};
  
  
          \draw[fill=white,thick, draw=black] (7,-.15) to [bend right=45] 
          ++(0,.15) to [bend left=45] ++(0,.15);
  
          \draw[fill=white,thick, draw=black] (.8,-\hlen-.15) to [bend right=45] 
          ++(0,.15) to [bend left=45] ++(0,.15);
          \end{scope}
      \end{scriptsize}
  \end{tikzpicture}


    \caption{(Left) Schematic of the different flow phenomena
      encountered over an aircraft challenging current CFD methodologies. 
      For compactness, all the phenomena are overlaid
      in the schematic; however, not all of them occur concurrently or
      even take place at all. (Top right) Number of wind tunnel tests
      required during the design cycle as a function of time. The
      horizontal axis also marks the introduction of widely used RANS
      models. The graph is adapted from~\cite{xiao2019}.
      (Bottom right) Selected milestones in wall and SGS modeling for (WM)LES of external aerodynamics. 
      The text in blue highlights milestones in 
      traditional (WM)LES; the text in red highlights machine-learning based 
      milestones in (WM)LES. In the figure, LoW stands for `law of the wall' and
      TBL for `thin boundary layer'. \label{fig:background} }
\end{figure}
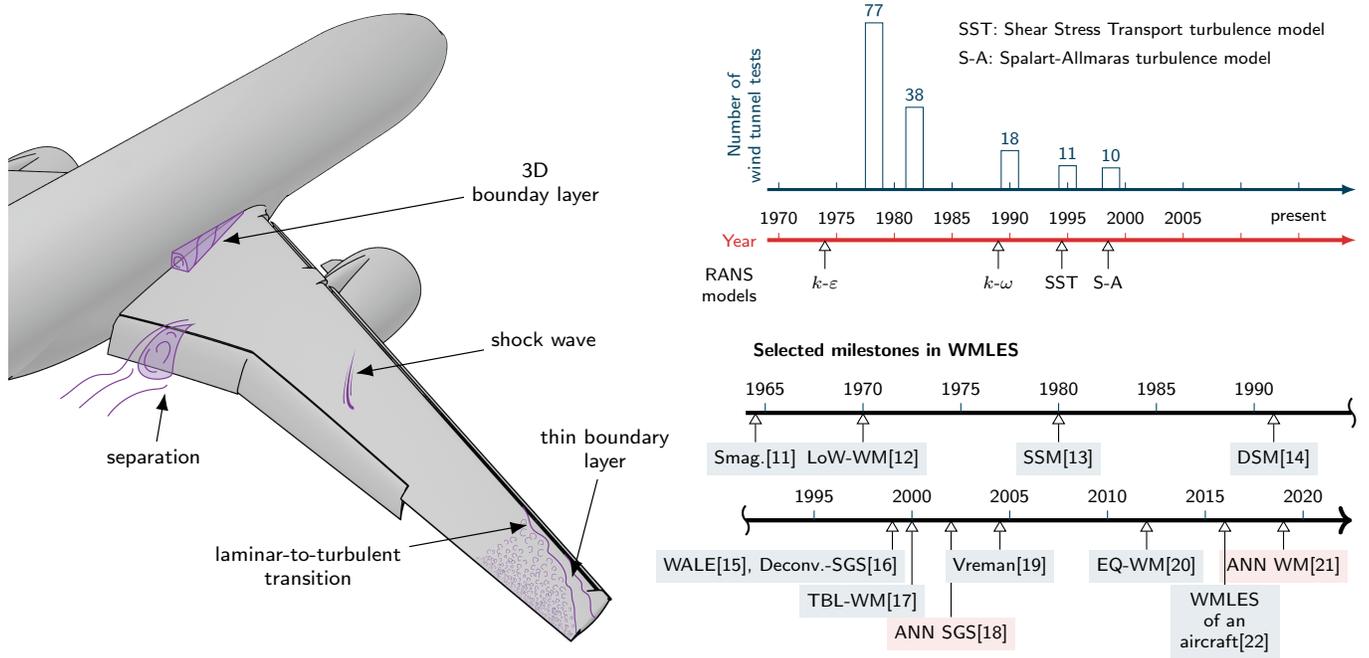

In terms of prediction and control, the field of CFD is in the
privileged position of having a highly accurate framework to model
flows: the Navier--Stokes equations. However, the number of degrees of
freedom involved in real-world applications ranges between $10^{13}$
to $10^{20}$, making the direct numerical simulation of all flow
scales infeasible. Consequently, the treatment of turbulence in
industrial CFD has primarily relied on closure models for the
Reynolds-averaged Navier-Stokes (RANS) equations~\cite{casey2000}. The
approach has materialized in various forms, from pure RANS solutions
to hybrid methods such as Detached Eddy Simulation and its
variants~\cite{spalart2009}. Many RANS models have been developed to
overcome the limitations of their predecessors, often by expanding and
calibrating their coefficients to account for missing
physics. However, no practical model is universally applicable across
the broad range of flow regimes of interest to the industry.  Examples
of challenging flow scenarios include separated flows,
favorable/adverse pressure gradient effects, shock waves, mean-flow
three-dimensionality, and transition to turbulence, among others, as
illustrated in Fig.~\ref{fig:background}(left).  In these situations,
RANS predictions tend to be inconsistent and unreliable, especially
for geometries and conditions representative of the flight envelope of
commercial airplanes~\cite{rumsey2018}. Further CFD experience with
aircraft at high angles of attack has revealed that RANS-based solvers
have difficulties predicting maximum lift and the physical mechanisms
for stall~\cite{kiris2022}. As a result, the number of wind tunnel
experiments required during the final stage of the aircraft design
cycle has remained at around ten for the past two decades
(Fig.~\ref{fig:background} top right).

Recently, large-eddy simulation (LES) has gained momentum as a tool
for both scientific investigations and industrial applications. In
LES, the large flow motions containing most of the energy are directly
resolved by the grid, while the dissipative effect of the small scales
is modeled through a subgrid-scale (SGS) model. By additionally
modeling the near-wall flow using a so-called wall model, the
grid-point requirements for wall-modeled LES (WMLES) scale at most
linearly with increasing Reynolds number~\cite{choi2012, yang2021,
  lozanoduran2022}, defined as the ratio between the largest and
smallest flow scales in the problem. Most widely used SGS models were
developed from the 1980s to the early 2000s and are based on the
eddy-viscosity assumption~\cite{smagorinsky1963} 
(see Fig.~\ref{fig:background} bottom right). Among the most
prominent SGS models, we can cite the similarity
model~\cite{bardina1980}, Vreman model~\cite{vreman2004}, dynamic
Smagorinsky model~\cite{germano1991}, deconvolution
model~\cite{stolz1999}, and sigma model~\cite{nicoud2011}, to name a
few. A detailed account of SGS models can be found in~\cite{sagaut2006}.  
Regarding wall modeling, several strategies
have been explored in the literature, and comprehensive reviews can be
found in refs.~\cite{cabot2000, piomelli2002, spalart2009,
  larsson2015, bose2018}. A widely adopted method for wall modeling is
the wall-flux approach. This approach substitutes the no-slip and
thermal boundary conditions at the wall with the shear stress and heat
flux values provided by the wall model. Popular examples of these
approaches include those based on the law of the
wall~\cite{deardorff1970, schumann1975, piomelli1989}, the
full/simplified RANS equations~\cite{balaras1996, wang2002, chung2009,
  bodart2011, kawai2012, park2014, yang2015}, vortex-based
models~\cite{pullin2000}, or dynamic wall models~\cite{bose2014,
  bae2019}.

In this context, machine learning (ML) has emerged as a powerful tool
to enhance existing turbulence modeling approaches in the fluid
community\cite{brunton2019, brenner2019, pandey2020, beck2021,
  duraisamy2021, vinuesa2022}. Over the last two decades, there have
been multiple efforts devoted to the development of SGS and wall
models using ML tools. Most ML-based SGS models rely on supervised
learning, which involves the discovery of a function that maps an
input to an output based on provided training input-output
pairs. Early approaches used artificial neural networks (ANNs) to
emulate and speed up conventional, yet computationally expensive, SGS
models\cite{sarghini2003}. More recently, SGS models have been trained
to predict the so-called perfect SGS terms using data from filtered
direct numerical simulation (DNS)\cite{gamahara2017, wang2018investigation,
zhou2019, xie2019, xie2020artificial,
  park2021, wang2021artificial, kang2023, kim2024, prakash2024}. 
  Other approaches involve deriving SGS terms from optimal
estimator theory\cite{vollant2017} and deconvolution
operators\cite{hickel2004, maulik2017, fukami2019, yuan2020, xie2020spatially}, as well as from
reinforcement learning\cite{novati2021, kim2022, kurz2023}.
Similar ML methods have been employed to devise wall models for LES
via supervised learning. One of the initial attempts can be found in
ref.~\cite{yang2019predictive}. Other examples include spanwise
rotating channel flows\cite{huang2019wall}, flows over periodic
hills\cite{zhou2021wall}, turbulent flows with
separation\cite{zangeneh2021data}, and boundary layer flows in the
presence of shock-boundary layer
interaction\cite{bhaskaran2021science}, with mixed results in
\emph{a posteriori} testing. Recent works have also leveraged semi-supervised
learning to devise wall models\cite{bae2022scientific}, although these
approaches are still limited to simple flow configurations.

Despite the progress made, ML-based SGS and wall models still face
challenges and have yet to serve as an effective solution for
addressing long-standing issues in turbulence modeling for CFD. Even
two decades after the introduction of the first ML-based SGS model for
LES\cite{sarghini2003}, there has been only a single demonstration of
an ML-based SGS model applied to a realistic aircraft
configuration. This demonstration was conducted using a prototype of
the model presented here\cite{ling2022}. The goal of this study is to
introduce a novel, unified SGS and wall model for WMLES, aiming to
bridge the gap between our current predictive capabilities and the
demanding requirements of the aerospace industry.


\section*{Results}

\subsection*{The building-block flow model}

The simulation of turbulent flows involves solving the equations for
conservation of momentum:
\begin{equation}
  \label{eq:NS}
  \frac{\partial \rho u_i}{\partial t} + \frac{\partial \rho u_i
    u_j}{\partial x_j} = -\frac{\partial p}{\partial x_i} +
  \frac{\partial \sigma_{ij}}{\partial x_j}, \ i=1,2,3,
\end{equation}
along with the conservation of mass, energy, and the equation of state
for the gas. Here, $\rho$ represents the flow density, $u_i$ denotes
the $i$-th velocity component, $p$ signifies the pressure, and
$\sigma_{ij}$ stands for the viscous stress tensor. Note that repeated
indices in Eq.~\eqref{eq:NS} imply summation.  The challenge arises in
numerically solving the equations on a computational grid with the
ability to capture the smallest scales of the problem. In most
realistic scenarios, this task becomes computationally intractable due
to the hundreds of billions of grid points required.  Typically,
affordable grids comprise of the order of hundred million
points. However, the latter grids fall short in capturing the smallest
flow motions that significantly contribute to the mean forces on the
vehicle. The solution lies in introducing a correction term to the
right-hand side of Eq.~\eqref{eq:NS}, denoted as $\partial
\tau^\mathrm{SGS}_{ij} /\partial x_j$, where $\tau^\mathrm{SGS}_{ij}$
is the SGS closure model. This corrective term accounts for the
physics of the small scales unresolved by the coarse grid. 

\begin{figure}
	\centering
    \ig[width=17.8cm]{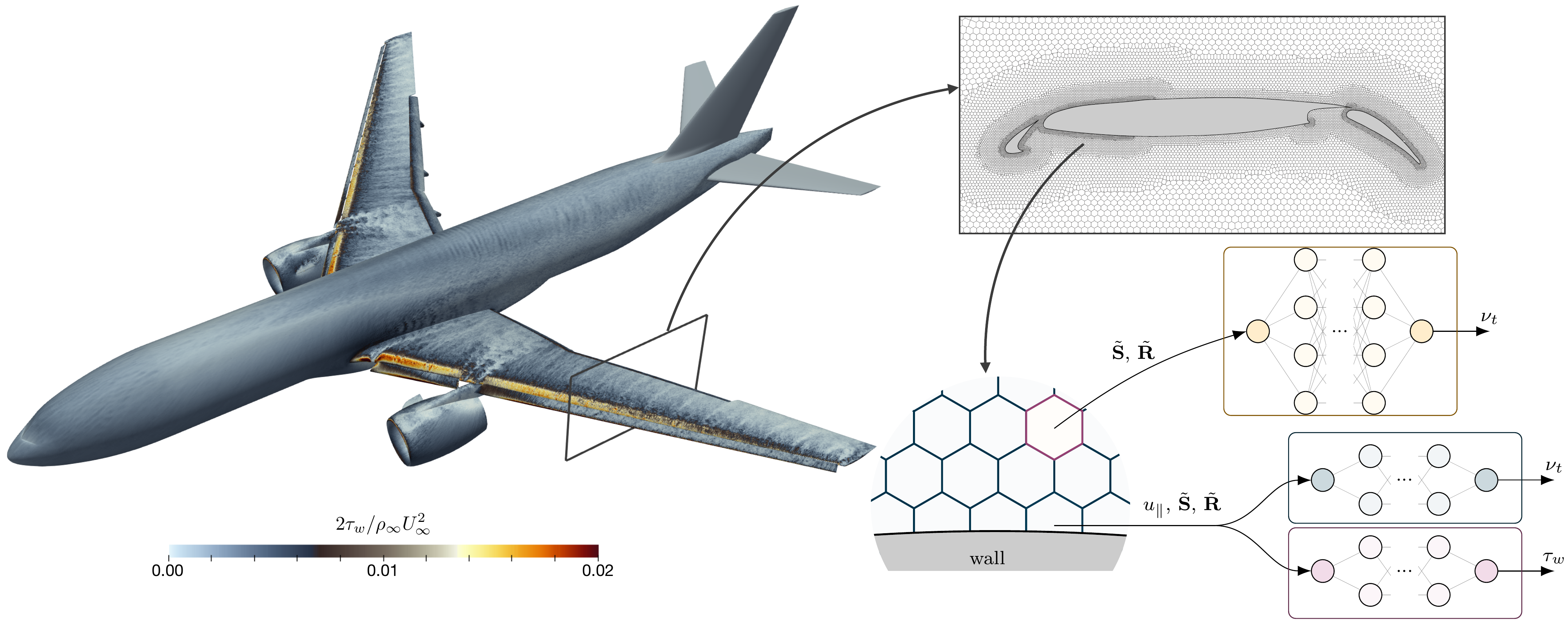}
	\caption{Schematic of the BFM applied to an aircraft in
          landing configuration. The tail of the aircraft is included
          in the model just for illustration purposes. The color over
          the surface of the aircraft represents the magnitude of the
          wall-shear stress normalized by the freestream velocity and
          density. The panel displays a cross-section of the grid
          close to the solid boundary, illustrating the three ANNs
          that constitute the BFM, along with their respective inputs
          and outputs. \label{fig:BFMa} }
\end{figure}

The Building-block Flow Model (BFM) provides the SGS tensor,
$\tau^\mathrm{SGS}_{ij}$ that tackles the challenge posed by missing
scales at both solid boundaries (traditionally addressed with a wall
model) and in the flow above these boundaries (traditionally modeled
with an SGS model). An overview of the model is shown in
Fig.~\ref{fig:BFMa}. The core modeling assumption is that the physics
of the missing scales in complex scenarios can be locally mapped onto
the small scales of simpler flows~\cite{ling2022, lozano2023}. Under
this premise, it is postulated that there is a finite set of simple
flows, referred to as building-block flows (BBFs), containing the
essential flow physics necessary to formulate generalizable SGS and
wall models.  In the following, we discuss the key components of the
BFM: BBFs, model architecture, and training data.
%
%
\subsection*{Building block flows}

\begin{figure}[t!]
    \centering
	\ig[width=1\tw]{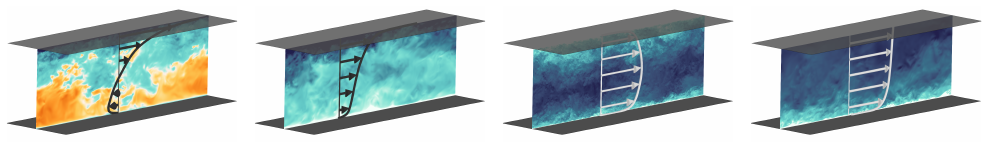}
	\caption{Examples of building-block flows (from left to
          right): wall-bounded turbulence under separation, adverse,
          zero, and favorable mean pressure gradient. The training
          database comprises several cases for each building block
          flow.  Dark blue corresponds to positive streamwise
          velocity, and orange corresponds to negative streamwise
          velocity. The solid lines and arrows represent the mean
          velocity profile. \label{fig:BFMb} }
\end{figure}

The BBFs offer simple representations of different flow regimes for
which scale-resolving, high-fidelity DNS data or experimental data are
available. The configuration of the BBFs entails an incompressible
turbulent flow confined between two parallel walls. Four types of BBFs
are considered: wall-bounded turbulence under separation, adverse,
zero, and favorable mean pressure gradient. Examples of these BBFs are
illustrated in Fig.~\ref{fig:BFMb}.  The intentional exclusion of
cases with additional complexities, such as airfoils, wings, bumps,
etc., is grounded in the idea that the BBFs should encapsulate
fundamental flow physics to predict more intricate scenarios. Hence,
the aim is to avoid case overfitting, e.g., correctly predicting the
flow over a wing merely because the model was trained on similar
wings. Note that the largest scales of the flow are intended to be
resolved by the computational grid, whereas the BBFs represent the
smaller flow features. This justifies the geometric simplicity of the
BBFs over more complex cases. Additionally, the computational
affordability of the BBFs allows for the collection of a large number
of high-fidelity data for training, which facilitates efficient
exploration of multiple flow regimes. This aspect is important as the
BFM must learn the physical scaling of the non-dimensional inputs and
outputs that control the problem at hand, rather than merely fitting
to a set of cases.

\subsection*{Model architecture}

The BFM is implemented using three feedforward ANNs (see
Fig.~\ref{fig:BFMa}).  The first ANN, adapted from the ML-based wall
model by Lozano-Dur\'an and Bae~\cite{lozano2023}, is tasked with
predicting the wall-shear stress at the solid boundaries
($\boldsymbol{\tau}_w$).  The second ANN predicts the SGS stress
tensor for control volumes adjacent to the solid boundaries, while the
third ANN does the same for the remaining control volumes. The ANNs
take as input the local values of the invariants of the symmetric
($\tilde{\mathbf{S}}$) and antisymmetric ($\tilde{\mathbf{R}}$) parts
of the velocity gradient tensor~\cite{lund1993}. Additionally, the
first and second ANNs incorporate information about the wall-parallel
velocity of the flow relative to the solid boundary ($u_{||}$). These
inputs are chosen to ensure invariance under constant translations and
rotations of the reference frame, along with Galilean invariance. The
output of the ANNs is the value of the SGS stress tensor via the eddy
viscosity $\nu_t$.  The inputs and output variables of the first ANN
are non-dimensionalized using viscous scaling (i.e., the kinematic
viscosity $\nu$ and the local grid size).  For the second and third
ANNs, the variables are non-dimensionalized using semi-viscous scaling
(based on $\nu$ and the first invariant itself).  The BFM is developed
for GPU architectures using the CUDA programming language, which is
particularly efficient in evaluating ANNs compared to the CPU
counterparts.

\subsection*{Numerically-consistent training data}

The training data is generated using a novel method that incorporates
an `exact-for-the-mean' SGS/wall model. The approach, referred to as
E-WMLES, involves conducting WMLES with a control mechanism that
identifies the required SGS tensor to predict specific mean statistics
of interest in the flow. The statistics considered here are the mean
velocity profiles and the mean wall-shear stress, which are obtained
from high-fidelity data of the BBFs. The BBFs were simulated using
E-WMLES in the same numerical flow solver and grid strategies
subsequently employed for implementing the BFM. The data generated
from these simulations were used to train the BFM, ensuring
consistency between the model and the numerical schemes of the flow
solver.

The E-WMLES approach diverges from the standard practice within the
community of training SGS models using filtered or coarse-grained
high-fidelity data. Our preference for the E-WMLES approach arises
from the known fact that the SGS tensor in WMLES is not consistent
with the filtered terms derived from the Navier-Stokes
equations~\cite{lund1995, lund2003, bae_brief_2017,
  bae_brief_2018}. This limitation is particularly significant in
external aerodynamics applications, where the typical grid sizes used
in WMLES are orders of magnitude ($10^2$ to $10^4$) coarser than the
smallest length scale of the flow. In these situations, numerical
errors become comparable to modeling errors, and both must be
accounted for by the SGS and wall model to yield accurate predictions.

\subsection*{Validation cases}

We evaluate the performance of the BFM in five cases: 
turbulent channel flows, a turbulent pipe flow, 
the separated flow over a bump, a simplified aircraft model 
and a realistic aircraft in landing configuration. 
The results are compared with simulations using
established SGS and wall models conducted using identical meshes as
those for the BFM. The models for comparison are the dynamic
Smagorisky SGS model (DSM)~\cite{germano1991, lilly1992} or Vreman
SGS model (VRE)~\cite{vreman2004} combined with the equilibrium wall
model (EQ)~\cite{kawai2012}. The models are labeled as DSM-EQ and
VRE-EQ, respectively. In the validation cases below, the computational
cost of conducting WMLES with BFM is approximately 0.9 times that of
DSM-EQ and 1.1 times that of VRE-EQ.

\subsection*{Turbulent channel flow}

\begin{figure}
\centering
\ig[width=\tw]{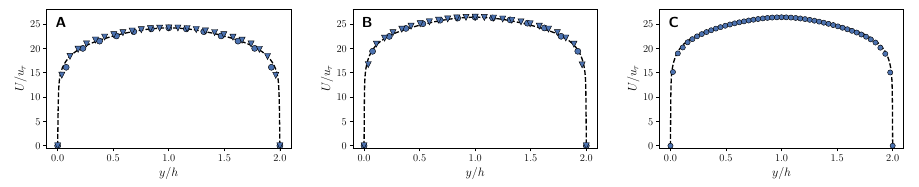}
\caption{The mean velocity profile in a turbulent channel flow as a
  function of the wall-normal distance at (A) Re$_\tau \approx 2000$
  for grid sizes $\Delta \approx 0.2h$, and $0.1h$; (B) Re$_\tau
  \approx 4200$ for grid sizes $\Delta \approx 0.2h$ and $0.1h$; and
  (C) Re$_\tau \approx 4200$ for $\Delta \approx 0.05h$.
    The dashed line is DNS data~\cite{lozano-duran2014} and the blue
    symbols are the BFM.
    In (A) and (B) circles correspond to $\Delta \approx 0.2h$ and
    triangles to $\Delta \approx 0.1h$.\label{fig:channel}}
\end{figure}

The first validation case is a turbulent channel flow, a canonical
configuration in which the flow is confined between two parallel
walls, as depicted by one of the BBFs in Fig.~\ref{fig:BFMb}.
Note that although this case is part of the training dataset, it
  is not guaranteed that BFM will perform well in \emph{a posteriori}
  testing due to potential inconsistencies between an ANN-based model
  and the numerical schemes of the
  solver\cite{beck2019,sirignano2020,macart2021,duraisamy2021}. 
  Therefore, this case serves as a demonstration that our strategy to enforce numerical consistency is successful in actual WMLES across various Reynolds numbers and grid resolutions. We also use this case to show the robustness of BFM to grid resolutions finer than those used for training.
  The
friction Reynolds numbers considered are Re$_\tau = u_\tau h/\nu
\approx 2000$ and $4200$, where $u_\tau$ is the friction velocity, $h$
is the channel half-height, and $\nu$ is the kinematic viscosity. The
channel is driven by a streamwise mean pressure gradient, such that Re$_c =
U_c h/\nu =48500$ and $112500$, respectively, with $U_c$ representing
the mean streamwise velocity at the centerline of the channel.  Three
spatially isotropic grid resolutions are tested: $\Delta = 0.05h$,
$0.1$ and $0.2h$, where the first was not included in the
  training dataset.
The BFM accurately predicts the mean velocity profile
(Fig.~\ref{fig:channel}) and the mean wall shear stress, achieving an
accuracy of 1\% to 6\% across the considered Reynolds numbers and grid
resolutions.

\subsection*{Turbulent pipe flow}
\providecommand{\cl}{\textit{c}}

\begin{figure}
  \centering
  \ig[width=.9\lw]{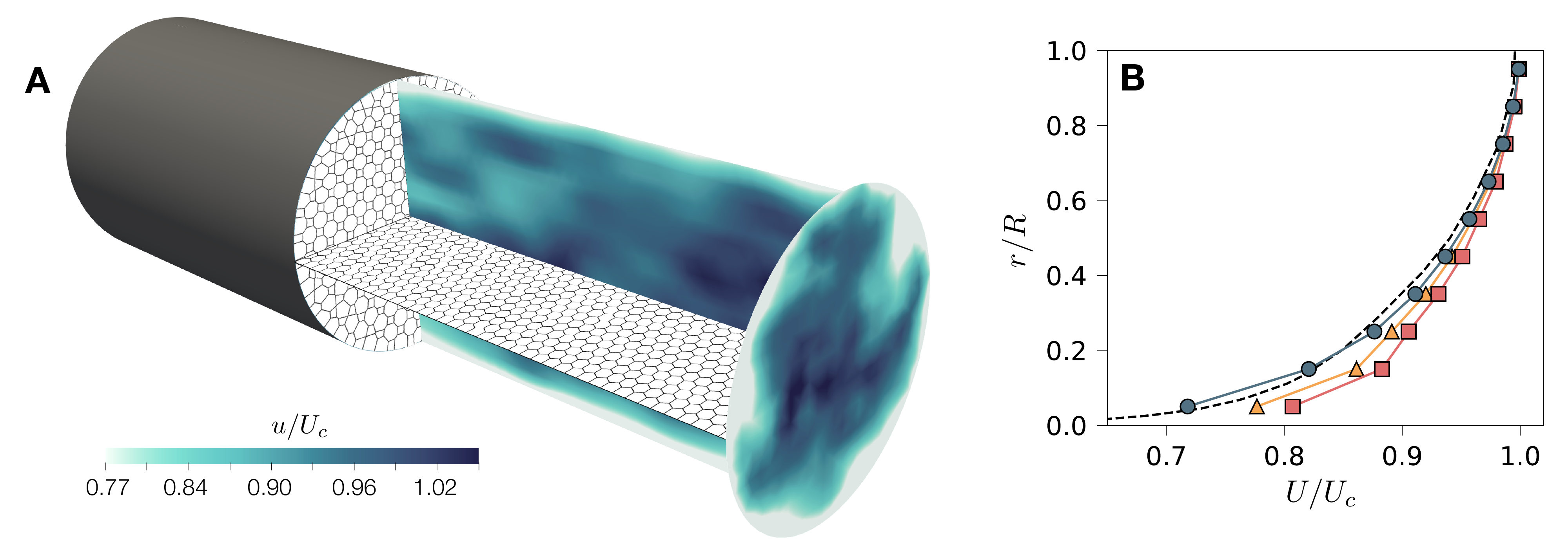}
  \caption{(A) Schematic and flow visualization of the pipe flow
      at $\Rey_\tau = 39500$. The radial and axial slices depict the
      grid structure and the instantaneous streamwise velocity. (B)
      Mean velocity profile along the radial direction $r$ from the
      wall ($r/R = 0$) to the pipe centerline ($r/R = 1$). Symbols
      correspond to BFM (blue circles), DSM-EQ (red squares), and
      VRE-EQ (yellow triangles). The dashed line is experimental
      values~\cite{baidya2019}.
  \label{fig:pipe}}
\end{figure}

The second validation case is the turbulent flow in a pipe (see
  Fig.~\ref{fig:pipe}A). We use this case to evaluate the performance
  of BFM at Reynolds numbers much higher than those it was trained
  for. The friction Reynolds number is $\Rey_\tau = u_\tau R / \nu =
  39,500$, where $R$ is the pipe radius, and the flow is
  incompressible. The pipe is assumed periodic in the streamwise
  direction with a length of $7.5R$. As in the turbulent channel flow,
  the pipe is driven by a streamwise mean pressure gradient such that
  $\Rey_{\cl} = U_\cl R / \nu \approx 1,272,800$, with $U_\cl$
  representing the mean streamwise velocity at the pipe
  centerline. The mesh is spatially isotropic (see Fig.~\ref{fig:pipe}A) 
  with a grid size of $\Delta = 0.1R$, which yields roughly 30,000 grid points.

We evaluate the performance of BFM, DSM-EQ, and VRE-EQ using the
  same flow conditions and computational grid. The results are
  compared with experimental
  data\cite{baidya2019}. Figure~\ref{fig:pipe}B displays the mean
  velocity profile, showing the improved agreement of BFM with the
  experimental results with respect to DSM-EQ and VRE-EQ. Similarly,
  BFM predicts the average wall shear stress within a 3\% error
  compared to the experiment, whereas DSM-EQ and VRE-EQ overpredict
  the average wall shear stress by 22\% and 13\%, respectively.

\subsection*{Separated flow over a  bump}\label{ssec:bump}

\begin{figure}[t!]
	\centering
	\ig[width=.9\lw,trim=0cm 0cm 0cm 0cm,clip]{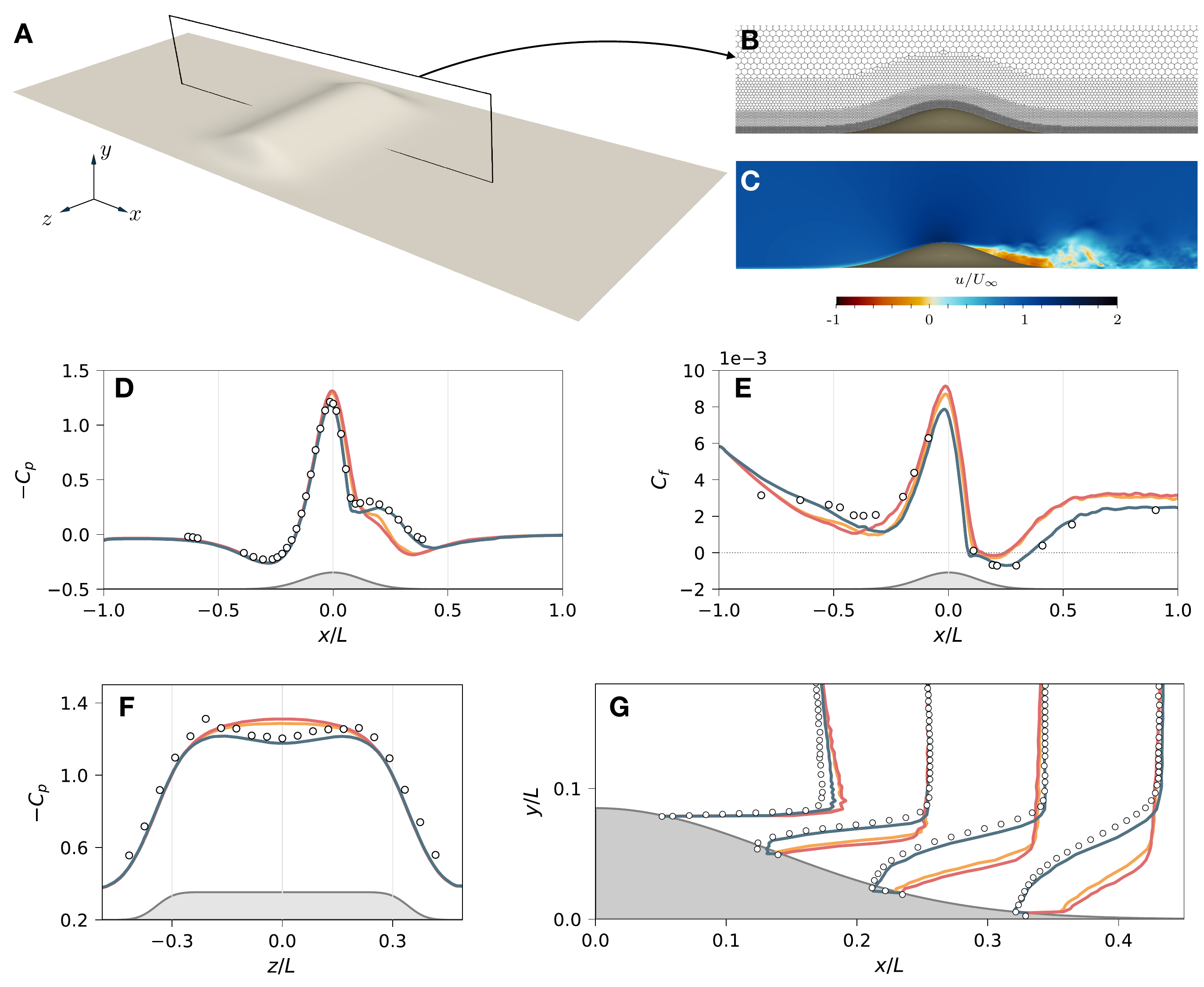}
    \caption{(A) Schematic of the Gaussian bump; (B) grid structure
      and (C) instantaneous streamwise velocity in the $y/L = 0$
      plane. The velocity of the freestream is $U_\infty$.  Note that
      the figure does not show the whole domain in the streamwise
      direction.  (D) Wall pressure coefficient $C_p = \frac{p -
        p_\infty}{\frac{1}{2} \rho_\infty U^2_\infty}$ and (E) wall
      friction coefficient $C_f = \frac{\tau_w}{\frac{1}{2}\rho_\infty
        U^2_\infty}$, along the $y/L = 0$ plane, where $U_\infty$,
      $p_\infty$ and $\rho_\infty$ are the freestream velocity,
      pressure and density, respectively, and $p$ and $\tau_w$ are
      averaged over time.  (F) Wall pressure coefficient along 
      the $x/L=0$ plane.
      (G) Mean velocity profile at $y/L=0$ at
      different streamwise locations.  In (D--G) The solid lines
      denote BFM (blue), DSM-EQ (red) and VRE-EQ (yellow). White
      circles are experimental
      values\cite{bumpTN,gray2022b}.\label{fig:bump}}
\end{figure}

The second validation case is a Gaussian bump designed by The Boeing
Company and the University of Washington~\cite{williams2020}, as shown
in Fig.~\ref{fig:bump}. The test consists of a three-dimensional
tapered hump at a Reynolds number of $3.6 \times 10^6$ and a Mach
number of $0.17$, based on the freestream quantities and the length of
the bump along the spanwise direction, $L$.
This test case has proven to be highly challenging for the RANS
  and LES community, with inaccurate prediction of the separation
bubble and non-monotonic convergence of the solution with grid
refinements\cite{iyer2021, agrawal2022} (see
  Fig.~\ref{fig:bumpsup}C--D).

\begin{figure}
  \centering
  \begin{tikzpicture}
    \node[anchor=south west] (fig) at (0,0) {\ig[width=1\tw,trim=0 .1cm 0
    0,clip]{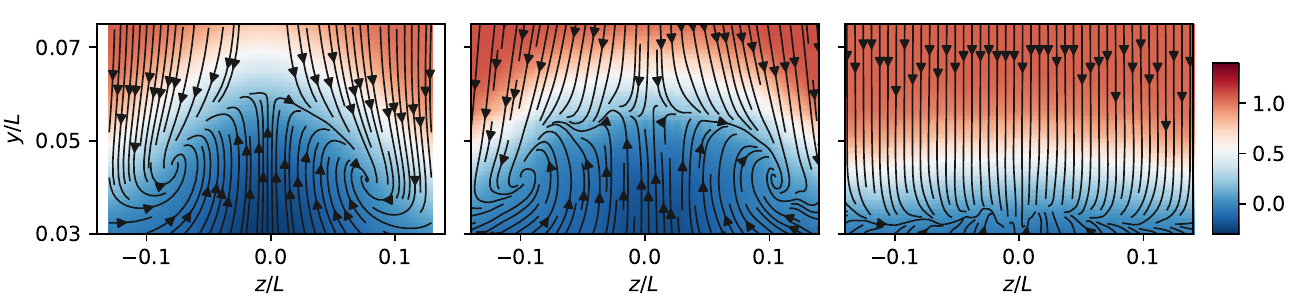}};
    \begin{scope}[x=(fig.south east),y=(fig.north west),font=\sffamily]
      \node[white] at (.105,.82) {\textbf{A}};
      \node[white] at (.38,.82) {\textbf{B}};
      \node[white] at (.67,.82) {\textbf{C}};
    \end{scope}
    \begin{scope}[x=(fig.south east),y=(fig.north west)]
      \node[black] at (.95,.85) {$U/U_\infty$};
    \end{scope}
  \end{tikzpicture}
  \caption{Mean streamwise velocity (colormap) and 
  in-plane streamlines at the plane $x/L = 0.208$. 
  (A) experiments\cite{bumpTN}, (B) BFM and (C) Vreman-EQ. 
  Colors range from (dark blue) $U = -0.3U_\infty$ to (dark red) 
  $U = 1.4U_\infty$. Note that axes are not to scale.
  \label{fig:bumpPIV}}
\end{figure}

We evaluate the performance of the BFM in predicting the mean velocity
profiles, and the mean pressure and friction at the wall
(characterized by the non-dimensional coefficients $C_p$ and $C_f$,
respectively). The simulation consists of roughly 9 million control
volumes. Similar WMLESes are conducted using DSM-EQ and VRE-EQ.  The
numerical findings are compared with experimental
data\cite{bumpTN}. Figure~\ref{fig:bump}(D,E) illustrates the
evolution of $C_p$ and $C_f$ along the centerline of the bump
($x/L$). Notably, the BFM yields more accurate predictions of the
$C_p$ and $C_f$ evolution in the separation region ($x/L>0.1$)
compared to DSM-EQ and VRE-EQ, which even fail to predict flow
separation. The improved predictions by the BFM also extend to the
presssure along the spanwise direction at the bump apex ($x/L =
  0$), as shown in Fig.~\ref{fig:bump}F, and the mean velocity
profiles, as demonstrated in Fig.~\ref{fig:bump}G. The BFM
predicts mean velocity with up to 3\% accuracy, whereas both DSM-EQ
and VRE-EQ yield a similar result that overestimate the mean
velocities by more than 30\%. 
Similarly, the BFM accurately predicts the flow topology of the separation. This is illustrated in Fig.~\ref{fig:bumpPIV}, which shows a reduced window (with limits $z/L = \pm0.13$ and $y/L \leq 0.075$) downstream of the bump. The BFM captures the counter-rotating vortices that lift away from the surface\cite{bumpTN}, in contrast to VRE-EQ, which predicts a completely different flow topology. The results for DSM-EQ (not shown) are similar to those of VRE-EQ.
Nevertheless, there is still room for
improvement upstream of the bump ($x/L<0$), where all models exhibit
low accuracy in the prediction of $C_f$ (Fig.~\ref{fig:bump}E).

\subsection*{Aircraft in landing configuration}

The third validation case is an aircraft in landing configuration,
specifically the NASA Common Research Model High-Lift (CRM-HL). This
case serves as the gold standard used by the aerospace community to
assess the capabilities of different CFD
methodologies~\cite{lacy2016development}. The CRM-HL is a
geometrically realistic aircraft that includes the bracketry
associated with deployed flaps and slats, as well as a flow-through
nacelle mounted on the underside of the wing
(Fig.~\ref{fig:BFMa}). The Reynolds number of the aircraft is 5.49
million based on the mean aerodynamic chord and freestream velocity,
and the freestream Mach number is 0.2.

The simulations are conducted using a grid with approximately 30
million control volumes.  Four angles of attack are considered:
$\alpha = 7.05^\circ$, $11.29^\circ$, $17.05^\circ$ and
$19.57^\circ$.  Figure~\ref{fig:BFMa} contains the model geometry, an
inset of the mesh, and a visualization of the wall-shear stress
predicted by BFM.

\begin{figure}
  \centering
	\ig[width=17.8cm]{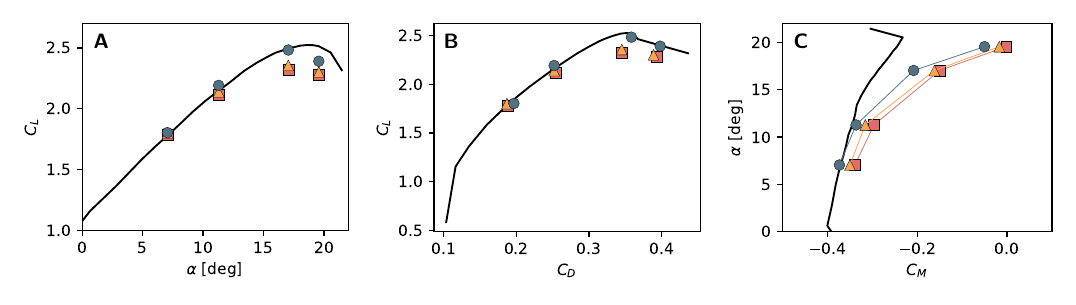}
	\ig[width=.95\tw]{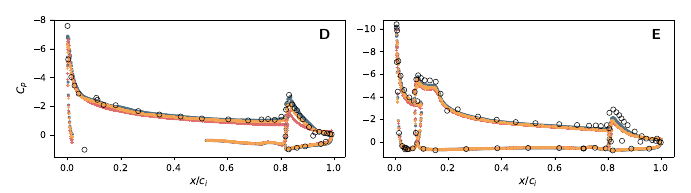}
	\caption{(A) Lift coefficient, (B) drag polar and (C) pitching
          moment coefficient for the CRM-HL. The black lines denote
          experimental results, blue circles are for the BFM, squares
          are for DSM-EQ and orange squared for VRE-EQ.
          Pressure coefficient along the chord of the wing at
            (D) the 33\% and (E) 55\% spanwise section of the wing for
            $\alpha = 17.05^\circ$. The chord location is normalized
            by the sectional chord-length $c_{\text{ref}}$. Empty
            circles represents experimental results, blue circles are
            for the BFM, orange and green plus symbols are for DSM-EQ
            and VRE-EQ, respectively.\label{fig:CRMforcescp}}
\end{figure}

Figure~\ref{fig:CRMforcescp}(A-C) displays the forces acting on the
aircraft, characterized by the lift ($C_L$), drag ($C_D$), and
pitching moment ($C_M$) coefficients for BFM, DSM-EQ, and VRE-EQ. The
results are compared with experimental data~\cite{evans2020test}.
The three plots should be interpreted as a whole, as recent
  studies\cite{rumsey2023} have demonstrated that --due to the
  complexity of the case-- accurate predictions of $C_L$ and $C_D$ may
  result from error cancellation during the integration of aerodynamic
  forces. This issue is less significant in the pitching moment, which
  is sensitive to force distribution and hence less susceptible to
  error cancellation. Overall, BFM offers similar or improved accuracy
  in predicting the three aerodynamic coefficients, especially at high
  angles of attack during stall and post-stall phases. It is important
  to note that despite the accurate predictions of lift and drag by
  DSM-EQ and VRE-EQ at $\alpha = 7.05^{\circ}$, these are accompanied
  by a poorer prediction of the pitching moment compared to BFM.

Inspection of the sectional pressure coefficient ($C_p$) along
  the wing, as depicted in Fig.~\ref{fig:CRMforcescp}(D,E) for two
  spanwise locations, reaffirms that BFM yields more accurate
  predictions compared to DSM-EQ and VRE-EQ. The improvements are more
  noticeable in the flaps (last 20\% of the wing chord).  The results
are still far from satisfactory, especially for the pitching moment,
and there is clear potential for further improvements. Nonetheless, it
is worth noting that BFM has never `seen' an aircraft-like flow or
been trained in a case resembling an airfoil or a wing. This
demonstrates the ability of the BFM methodology to offer predictions
that go beyond simple regression.

\subsection*{Juncture flow in simplified aircraft}

\begin{figure}
    \centering
        \ig[width=\tw,trim=0 8cm 0 0cm,clip]{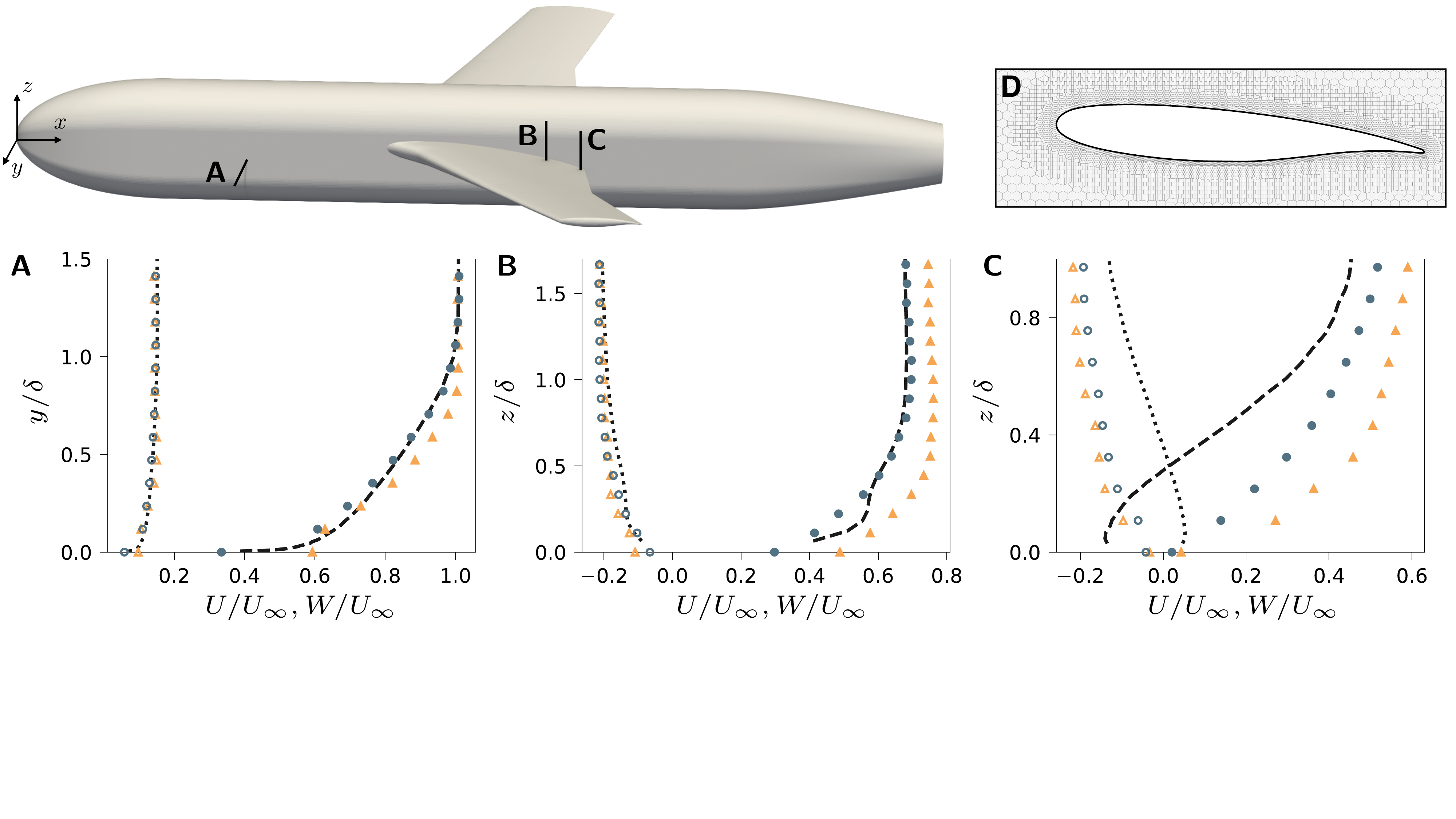}
        \caption{Mean velocity profile in the NASA Juncture Flow
            Experiment. The probes are located at (A) the upstream
            region of the fuselage ($x=1168.4$ mm, $z=0$ mm), (B) the
            wing-body junction ($x=2747.6$ mm; $y=239.1$ mm), and (C)
            the wing-body junction close to the trailing edge
            ($x=2922.6$ mm; $y=239.1$ mm). The approximate location
            are depicted in the sketch at the top.  In all the plots,
            the symbols denote: BFM (blue circles) and VRE-EQ (yellow
            triangles). Closed symbols corresponds to $U/U_\infty$
            and open symbols to $W/U_\infty$, where $U_\infty$ is the
            freestream velocity.  The lines correspond to experimental
            values\cite{rumsey2018}. $\delta$ is the local boundary layer 
            thickness~\cite{lozanoduran2022}. (D) Cross-section of the grid at
            the wing.\label{fig:juncture}}
\end{figure}

We evaluate the performance of the BFM using data from the NASA
  Juncture Flow Experiment\cite{rumsey2018}. This experiment involves
  a simplified aircraft configuration, as illustrated in
  Fig.~\ref{fig:juncture}, featuring flow separation at the juncture
  between the fuselage and the trailing edge of the
  wing\cite{rumsey2018}. The Reynolds number is calculated based on
  the fuselage length ($L = 4839.2$ mm) and the free-stream velocity,
  yielding $\Rey = 20.8 \times 10^6$. Additionally, the free-stream
  Mach number is $M = 0.189$, and the angle of attack is $\alpha =
  5^\circ$. A cross-section of the grid, containing approximately 30
  million points, is depicted in Fig.~\ref{fig:juncture}D. This
  analysis complements the CRM-HL case previously discussed, as the
  experimental campaign of the Juncture Flow contains mean velocity profiles at various
  locations, which were not available for the CRM-HL. Unlike
  integrated measurements such as lift and drag coefficients, mean
  velocity profiles are less susceptible to error cancellation, thus
  facilitating a more detailed evaluation of the models.

WMLES was conducted using both BFM and VRE-EQ.  The results for the mean streamwise ($U$) and spanwise ($W$) velocity components are compared with experimental measurements at three different locations, as shown in Figure~\ref{fig:juncture}(A--C). The symbols in the figure represent the actual locations of the grid points. BFM yields accurate results in sections A and B, showing an improvement over VRE-EQ. BFM also improves predictions at section C. However, both models fail to capture the flow separation. This failure is not surprising, given that the thickness of the separation region is represented by only two grid points in the wall-normal direction, making accurate prediction extremely challenging. Despite this limitation, BFM still provides enhanced predictions of the mean velocity profile compared to VRE-EQ across the same grid resolution at all three locations. WMLES was also performed using DSM-EQ (not shown), which yielded slightly poorer predictions than VRE-EQ.

\section*{Discussion}
\label{sec:discussion}

\subsection*{Future opportunities of BFM for WMLES}

The BFM has the potential to significantly advance the resolution of
key Grand Challenges in the aerospace industry.  These challenges
include simulating an aircraft across its entire flight envelope and
addressing off-design turbofan engine transients --both highlighted in
the NASA CFD Vision 2030 report~\cite{slotnick2014}. The BFM can also
play a significant role in Certification by Analysis, i.e., the
development of simulation-based methods of compliance for airplane and
engine certification~\cite{mauery2021}. This marks a long-awaited
milestone in the aerospace industry, with the prospect of
Certification by Analysis estimated to result in substantial cost
savings, amounting to billions of dollars in the design process.

A distinctive advantage of the BFM, which might contribute to
addressing the aforementioned challenges compared to previous models,
is its ability to accommodate various flow regimes through the BBF
database. The current version of BFM includes separated flow and the
effects of zero, adverse, and favorable mean pressure
gradients. However, the process of incorporating additional flow
physics is scalable by expanding the building block database, a
capability absent in current SGS/wall models. 
The impact of excluding BBFs from the BFM training set is illustrated in the Methods section for the Gaussian Bump.
Future versions of BFM
could involve BBFs accounting for laminar flow, compressibility
effects, shock waves, wall roughness, and the laminar-to-turbulent
transition. Efforts to include these new physics are already underway
in our group.

Finally, the machine-learning nature of the BFM also unlocks other
opportunities for uncertainty quantification and adaptive grid
refinement. These might be accomplished by computing a confidence
score based on the similarity between the actual input data and the
BFM training data. This feature has already been tested in a
preliminary version of the BFM, which incorporates only a wall
model\cite{lozano2023}. The work has demonstrated that a confidence
score can be determined by the distance of the input data to the
closest sample in the training set. If the input data appears unusual,
the model assigns a low confidence score, indicating that the flow is
unfamiliar and does not align with any knowledge from the
database. Regions with low confidence scores can be used to assist in
automatic grid refinement in subsequent simulations, to identify the
need for additional building blocks, and for uncertainty
quantification.

\subsection*{Conclusions}

We have introduced the Building-Block Flow Model (BFM) for large-eddy
simulation. The model is designed to address challenges encountered in
CFD within the aerospace industry, specifically the demand for
accurate and robust solutions at an affordable computational cost. The
core assumption of the BFM is that the subgrid-scale physics of
complex flows can be effectively represented by the physics of simpler
canonical flows. Implemented using three feedforward ANNs for GPU
architectures, the BFM is applicable to arbitrarily complex
geometries.  Unlike previous models, the training data for the BFM is
derived from controlled WMLES with an `exact' model for the mean
quantities of interest. This approach ensures consistency with the
numerical discretization and grid structure of the solver. We have
shown that, at the grid resolutions considered here, the BFM matches or
improves the predictions from conventional SGS and wall models in
both simple and complex scenarios.

Truly revolutionary improvements in WMLES will encompass advancements
in numerics, grid generation, and wall/SGS modeling. The BFM addresses
all these aspects by devising SGS/wall models consistent with the
numerics and the grid. The modularity of the BFM opens up new
opportunities for developing SGS/wall models that have broad
applicability across different scenarios. In essence, the BFM offers a
single model that can accurately represent various flow physics,
eliminating the need for multiple specialized models for specific flow
types.  To enhance model performance, we will continue the training of
future versions of the BFM with additional building blocks to account
for a richer collection of flow physics and grid resolutions to improve its robustness.


\section*{Methods}

\subsection*{Numerical solver and conventional models}

The BFM provides the closure model for the SGS stress tensor
$\tau_{ij}^\mathrm{SGS}$ for the compressible LES equations
\begin{eqnarray}
	\frac{\partial \bar{\rho}}{\partial t} + \frac{\partial \bar{\rho} \tilde{u}_i }{\partial x_i}  &=& 0, \label{eq:LES1}\\
	\frac{\partial \bar{\rho} \tilde{u}_i}{\partial t} + \frac{\partial \bar{\rho} \tilde{u}_i \tilde{u}_j}{\partial x_j} &=&
	-\frac{\partial \bar p}{\partial x_i} +  \frac{\partial \tilde{\sigma}_{ij}}{\partial x_j} - \frac{\partial \tau^\mathrm{SGS}_{ij}}{\
		\partial x_j}, \label{eq:LES2}\\
	C_v \frac{\partial \bar{\rho} \tilde{T}}{\partial t} +  C_v\frac{\partial \bar{\rho} \tilde{u}_j \tilde{T}}{\partial x_j} &=&
	-\bar p \frac{\partial \tilde{u}_j}{\partial x_j} + \tilde{\sigma}_{ij} \frac{\partial \tilde{u}_j}{\partial x_i}
    \\ &+& \frac{\partial}{\partial x_j} \left( \tilde{\kappa}  \frac{\partial \tilde{T}}{\partial x_j} \right) -
	C_v  \frac{\partial q^\mathrm{SGS}_j}{\partial x_j}, \label{eq:LES3}
\end{eqnarray}
where repeated indices imply summation, $(\bar{\cdot})$ denotes
filtered quantities, $(\tilde{\cdot})$ is the Favre average, $u_i$ is
the $i$-th velocity component, $\rho$ is the density, $T$ is the
temperature, $\kappa$ is the thermal conductivity, $C_v$ is the
specific heat at constant volume, and $\sigma_{ij}$ is the viscous
stress tensor. Note that repeated indices imply summation.  The heat
flux is computed as $q^\mathrm{SGS}_j = (\bar{\rho} \nu_t
/\text{Pr})\partial \tilde{T}/ \partial x_j$, where Pr is the
turbulent Prandtl number and $\nu_t$ is the eddy viscosity.  The
applications considered in this work are low speed flow, and heat
transfer and compressibility effects play no significant role.

The simulations are conducted with the solver GPU version of charLES
developed by Cascade Technologies,
Inc. (Cadence)~\cite{bres2018}. The code integrates the compressible
LES equations using a kinetic-energy conserving, second-order
accurate, finite volume method.  The numerical discretization relies
on a flux formulation which is approximately entropy preserving in the
inviscid limit, thereby limiting the amount of numerical dissipation
added into the calculation. The time integration is performed with a
third-order Runge-Kutta explicit method. The ANNs of the BFM are
evaluated using the GPU capabilities of the solver.

The mesh generation follows a Voronoi hexagonal close-packed
point-seeding method, which automatically builds locally isotropic
meshes for arbitrarily complex geometries. First, the water-tight
surface geometry is provided to describe the computational
domain. Second, the coarsest grid resolution in the domain is set to
uniformly seeded points. Additional refinement levels are specified in
the vicinity of the walls if needed.  Thirty iterations of Lloyd's
algorithm are undertaken to smooth the transition between layers with
different grid resolutions.

The dynamic Smagorinsky model~\cite{germano1991} with the
modification by Lilly~\cite{lilly1992} and the Vreman
model~\cite{vreman2004} are considered as SGS models.  For the wall,
we use an equilibrium wall model.  The no-slip boundary condition at
the wall is replaced by a wall-stress boundary condition.  The walls
are assumed adiabatic and the wall stress is obtained by an algebraic
equilibrium wall model derived from the integration of the
one-dimensional stress model along the wall-normal
direction~\cite{piomelli1989},
\begin{equation}
	u_{||}^+(y_\perp^+) =
	\begin{cases}
		y_\perp^+ + a_1 (y_\perp^{+})^2 \text{\, \, \, for $y_\perp^+ < 23$}, \\
		\frac{1}{\kappa}\ln{y_\perp^+} + B \text{\, \, \, \, otherwise}
	\end{cases}
	\label{eq:charles_algwm}
\end{equation}
where $u_{||}$ is the model wall-parallel velocity at the second grid
point off the wall, $y_\perp$ is the wall-normal direction to the
surface, $\kappa=0.41$ is the K\'arm\'an constant, $B = 5.2$ is the
intercept constant, and $a_1$ is computed to ensure $C^1$
continuity. The superscript $+$ denotes inner units defined in terms
of wall friction velocity and the kinematic viscosity.

\subsection*{BFM formulation}

The anisotropic component of the SGS stress tensor is given by the
eddy-viscosity model
\begin{equation}
	\tau^\text{SGS}_{ij} = -2\nu_t \tilde{S}_{ij},
	\label{eq:eddy}
\end{equation}
where $\tilde{S}_{ij}$ is the rate-of-strain tensor.  The eddy
viscosity is a function of the first five invariants ($I_k$) of
rate-of-strain ($\mathbf{\tilde S}\equiv \tilde S_{ij}$) and
rate-of-rotation tensors ($\mathbf{\tilde R} \equiv \tilde
R_{ij}$)~\cite{lund1993},
\begin{equation}
	\nu_t = f(I_1,I_2,I_3,I_4,I_5, \theta),
	\label{eq:map}
\end{equation}
where $f$ represents a feedforward ANN, and $\theta$ denotes
additional input variables, namely, $\nu$, $\Delta$ and
$u_{\parallel}$, where $\nu$ is the kinematic viscosity, $\Delta$ is
the characteristic grid size and $u_{\parallel}$ is the magnitude of
the wall-parallel velocity measured with respect to the wall. The
latter is only used by the ANNs acting on the control volumes adjacent
to the walls. The invariants are defined as
\begin{equation}
	\begin{array}{ccccc}
        I_1=\operatorname{tr}\left(\mathbf{\tilde {S}}^2\right), & 
        I_2=\operatorname{tr}\left(\mathbf{\tilde {R}}^2\right), & 
        I_3=\operatorname{tr}\left(\mathbf{\tilde {S}}^3\right),\\
        I_4=\operatorname{tr}\left(\mathbf{\tilde {S} \tilde {R}}{ }^2\right), &
        I_5=\operatorname{tr}\left(\mathbf{\tilde {S}}^2 \mathbf{\tilde {R}}^2\right).
	\end{array}
\end{equation}

The ANNs are fully-connected feedforward networks.  The first ANN,
tasked with predicting the wall-shear stress, consists of 6 layers
with 40 neurons per layer. The second near-wall ANN consists of 10
layers with 12 neurons per each layer; and the outer-layer ANN
consists of 10 layers with 16 neurons per layer.  
The rationale
  for dividing the model into three ANNs is twofold: 1) task division
  (wall model versus SGS stress model), and 2) compatibility
  constraints (between the wall model and the near-wall SGS stress
  model). For the former, the initial ANN is responsible for
  predicting the wall shear stress (acting as a wall model), while the
  other two ANNs predict the eddy viscosity $\nu_t$. The second point
  relates to the ANN calculating $\nu_t$ at the control volumes
  adjacent to the wall. This ANN is not only tasked with obtaining
  $\nu_t$ for accurate mean flow predictions but also with ensuring
  compatibility with the wall model predictions, as the latter are
  tightly coupled with $\nu_t$ just above the boundary. A consequence
  of this coupling is that the output from the near-wall ANN scales
  differently from that of the ANN dedicated to the rest of the
  flow. Although it is feasible to train one single ANN for $\nu_t$,
  it was found that a better-performing model could be trained by
  separating the ANN into two.

The non-dimensionalization of the input and output features is
attained by using parameters that are local in both time and space to
guarantee the applicability of the model to complex geometries\cite{prakash2024}.
For the ANN responsible for predicting the wall-shear stress, the input
and output quantities are non-dimensionalized using viscous scaling:
$\nu$ and $\Delta$.  For the ANNs predicting the eddy viscosity, the
input and output variables are non-dimensionalized using semi-viscous
scaling: $(\nu\sqrt{\mathbf{\tilde S}:\mathbf{\tilde S}})^{1 / 2}$ and
$\Delta$, where the former represents a local velocity scale
  akin to the wall friction velocity $u_\tau$ typically used in
  wall-bounded flows, and the use of $\Delta$ as a length scale allows
  for accommodating the effect of different grid resolutions.

\subsection*{Training data preparation}

The mean velocity profiles and wall-shear stress are extracted from
DNS data of the BBFs. The turbulent channel flows are used to model
the regime where turbulence is fully developed without significant
mean-pressure gradient effects. The core region of the turbulent
  channel, where mean shear effects are weak, serves to model the SGS
  physics of isotropic turbulence. Separation, adverse and favorable
mean pressure gradient effects are learned from turbulent
Poiseuille-Couette (TPC) flows. In these cases, the top wall moves at
a constant speed ($U_t$) in the streamwise direction, and a mean
pressure gradient is applied in the direction opposed to the top wall
velocity. The pressure gradient ranges from mild to strong, so that
the flow ``separates'' (i.e., zero wall shear stress) on the bottom
wall. Favorable mean pressure gradient effects are obtained from the
upper wall of the TPC cases.

For turbulent channel flows, the mean DNS quantities are obtained from
the database in refs.\cite{lozano2014effect,hoyas2022}. Five cases
with friction Reynolds numbers 550, 950, 2000, 4200 and 10000 are
used.  For the TPC flows, the data was obtained from \cite{lozano2023}.  
The Reynolds numbers based on the mean pressure gradient are
$\mathit{Re_{P}} = {\sqrt{h^3\mathrm{d}P/\mathrm{d}x}}/\nu= 340, 680$
and $962$, and the Reynolds number based on the top wall velocity is
$\mathit{Re_U} = U_t h /\nu=22,360$, where $h$ is the channel
half-height.  The computational domain in the streamwise, wall-normal,
and spanwise direction is $L_x\times L_y \times L_z=4\pi
h\times2h\times 2\pi h$ for channel flows and $L_x\times L_y \times
L_z=2\pi h\times2h\times \pi h$ for TPC cases.

To generate the training data, we perform WMLES simulations of the
BBFs adjusting $\nu_t$ to match the mean DNS velocity profile and
correct wall-shear stress using the E-WMLES
approach\cite{lozano2023}. The simulations are performed in charLES.
The eddy viscosity is adjusted by finding a correcting factor $k(y)$
to a base model $\nu^{\text{base}}_t$ (in this case Vreman SGS
model~\cite{vreman2004}),
\begin{equation}\label{eq:nucorr}
	\quad \nu_t(x,y,z,t) = \nu^{\text{base}}_t(x,y,z,t) k(y),
\end{equation}
where $x$, $y$, and $z$ are the streamwise, wall-normal, and spanwise
directions, respectively, of the BBFs.  The correcting
factor is in turn computed by solving the optimization problem
\begin{equation}\label{eq:opt}
	\arg \min_{k(y)} \quad \int | U^{\text{DNS}}(y) - \tilde U(y) |^2 \mathrm{d}y
\end{equation}
where $U^{\text{DNS}}$ is the mean velocity profile from DNS and
$\tilde U$ is the mean velocity profile from WMLES obtained using the
eddy viscosity in \eqref{eq:nucorr}.  The free Conjugate-Gradient
algorithm and the Bayesian Global optimization algorithm~\cite{bayopt}
are used to minimize \eqref{eq:opt} for the turbulent channel and the
TPC flows, respectively. A Dirichlet non-slip boundary condition is
applied at the walls and the correct wall shear stress is enforced by
augmenting the eddy viscosity at the walls such that
$\left.\nu_t\right|_w=\left.\left(\partial \tilde {u}/\partial
y\right)\right|_w ^{-1} \tau_w/\rho-\nu$, following
\cite{bae2021effect}, where the subscript $w$ denotes quantities at
the wall, and $\tau_w$ is the mean wall-shear stress.  The grid size
for the WMLES cases with E-WMLES is $\Delta \approx 0.1h$ and $\approx
0.2h$ for the turbulent channel flows and $\Delta \approx 0.1h$ for
the TPC flows.

The optimization process for a given case is as follows: 1) WMLES
simulation is performed with a fixed $\tau_w$ --equal to the correct
value from DNS simulations, $\tau_w^{\text{DNS}}$-- and with an
initial random $k(y)$; 2) the simulation is run until the statistical
steady state is reached; 3) the integral in \eqref{eq:opt} is
evaluated, and 4) a new guess of $k(y)$ is provided by the optimizer.
This approach is continued until the condition
\begin{equation*}
	\frac{| U^{\text{DNS}}(y) - \tilde U(y) |}{U^{\text{DNS}}(y)} < 0.03
\end{equation*}
at each $y$ location for the turbulent channel flow cases is
satisfied.  For the TPC cases, this condition was too stringent, since
velocities are close to $0$ near the wall, and we relaxed the
condition to
\begin{equation*}
	\frac{| U^{\text{DNS}}(y) - \tilde U(y)|}{U_t} < 0.02.
\end{equation*}
Other approaches have been proposed in the literature to enforce
  compatibility between models and numerical schemes, most of them in
  the context of the RANS equations \cite{duraisamy2021,zhang2022ensemble}. 
  Advances in
  model-consistent approaches for LES are more limited due to the
  chaotic nature of the system and the inherent inconsistency between
  the filtered Navier-Stokes and the LES equations. Noteworthy
  examples include reinforcement learning \cite{novati2021,
    bae2022, zhou2022}, filter-consistent formulations
  \cite{bae_brief_2017, bae_brief_2018}, and optimal LES formulations
  \cite{langford2004, zandonade2004, lozano2023}.

\subsection*{Gaussian bump: details of the computational set-up}

The computational domain and set-up is as in~\cite{agrawal2022}.
The domain is a rectangular prism that extends from $-L$ to $1.5L$ in
the streamwise direction (with respect to the bump apex), $\pm0.5L$ in
the spanwise direction, and from $0L$ to $0.5L$ in the vertical
direction.
The lateral and top boundaries are free-slip, a constant uniform
inflow is imposed at the inlet, and the non-reflecting characteristic
boundary condition with constant pressure is applied at the outlet.

For the results presented in Fig.~\ref{fig:bump}, the grid has 
three levels of isotropic refinement.
Each refinement level has roughly 10 control volumes along the
wall-normal direction and the average size of each level is twice the
size of the previous level (see Fig.~\ref{fig:bump}).
The grid size of the layer closer to the wall is the smallest with
$\Delta_{\min} \approx 0.026h$, where $h = 0.0838L$ is the bump
height.
The number of control volumes is 8.7 million.  
This resolution yields of the order of 5 points per boundary layer at
the bump apex.
The simulations have been performed with a varying time step to ensure
that the Courant-Friedrichs-Lewy number is less than 2.

\subsubsection*{Effect of the grid size}

The current version of BFM has been trained on coarse grids with
  5 to 10 points per boundary layer. The behavior of BFM when
  extrapolating on finer grids is shown in
  Figure~\ref{fig:bumpsup}(A,B) for the separated flow over the bump.
  We consider a finer grid with a minimum grid size
  ($\Delta_{\min}~\approx~0.015h$) that is roughly half the one
  presented in the Results, leading to a mesh with 29 million control
  volumes.  The prediction of BFM becomes slightly less accurate
  upstream the bump, with lower $C_f$ and $-C_p$ predictions at the
  bump apex; however the prediction downstream the apex are consistent
  with the experiments.  We note that VRE-EQ and DSM-EQ still fails to
  predict the separation.

\subsubsection*{Comparison with RANS and WMLES}

Figure~\ref{fig:bumpsup}(C,D) displays the $C_p$ and $C_f$
values obtained using DSM-EQ and VRE-EQ on grids with 452 million
control volumes ($\Delta_{\min} \approx 0.003h$), as reported in
ref.~\cite{agrawal2022}. The figure also includes results from RANS
simulations using the Spalart-Allmaras (S-A) model\cite{SAmodel} and
the (S-A)-RC-QCR2000\cite{spalart2000}, performed by Iyer \&
Malik\cite{iyer2021} on a grid with 21 million cells (simulating
only the half span), which is effectively five times the grid size
used by BFM. While all simulations show convergence upstream of the
bump, VRE-EQ and S-A fail to capture the flow separation; the
(S-A)-RC-QCR2000 predicts separation but inaccurately in terms of
$C_p$ within the bubble and $C_f$ downstream. Only DSM-EQ with 452
million control volumes follows closely with the experimental
results. Remarkably, comparable accuracy was achieved by BFM on a
grid with fewer than 9 million control volumes, which contains 52
times fewer control volumes.

\subsubsection*{Effect of the building-blocks flows on the BFM accuracy}

\begin{figure}
  \centering
  \ig[width=.9\lw,trim=0 2cm 0 0cm,clip]{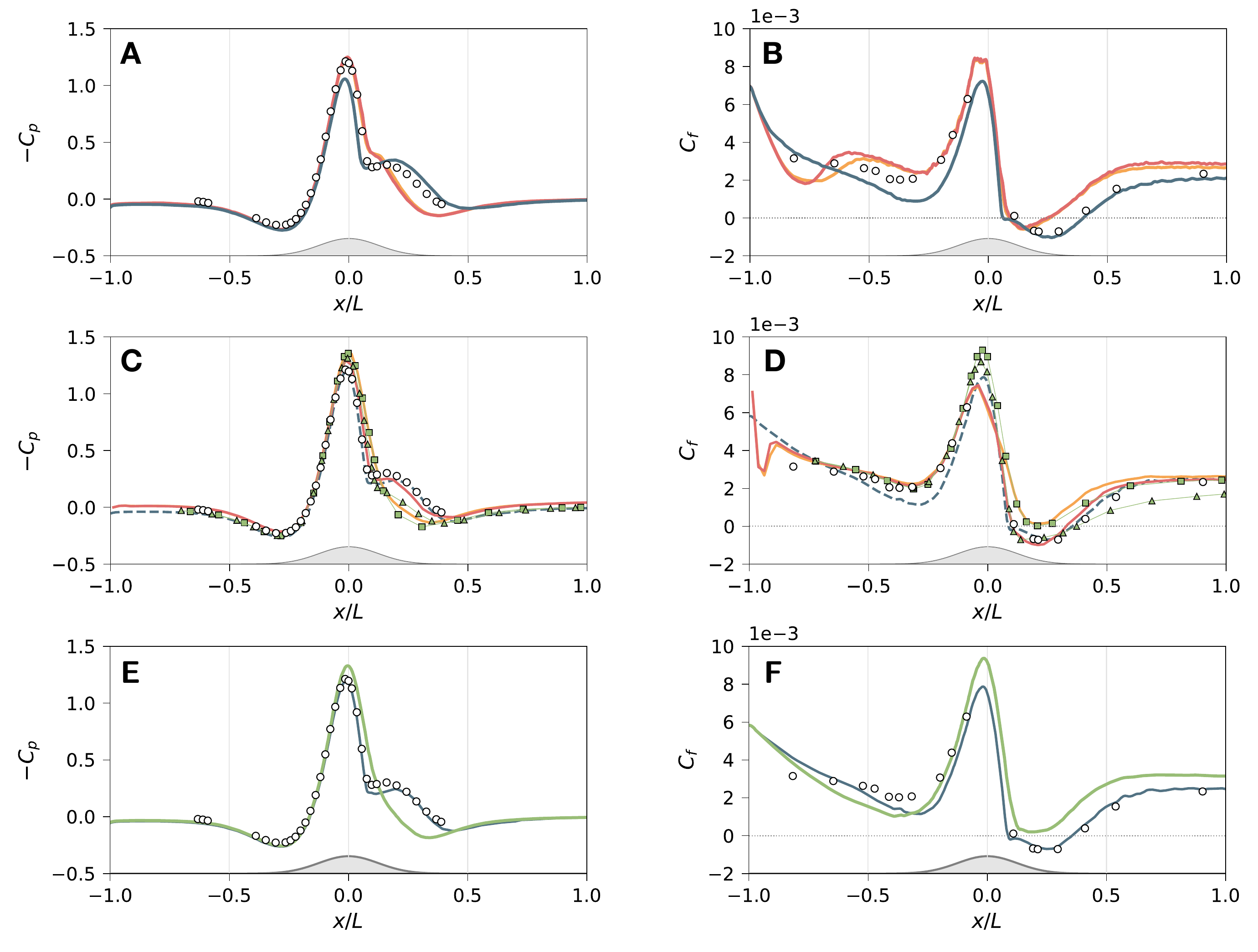}
  \caption{(A,C,E) Wall pressure and (B,D,F) wall friction
      coefficients along the $y/L = 0$ plane.  In (A,B), the results
      are obtained in a grid with $\Delta_{\min} \approx
      0.015h$. Legend as in Fig.~\ref{fig:bump}.  In (C,D), the lines
      correspond to DSM-EQ (red) and VRE-EQ (yellow) in a grid with
      $\Delta_{\min} \approx 0.003h$ ($\approx$450 million control
      volumes)\cite{agrawal2022}. The symbols are S-A (squares),
      (S-A)-RC-QRC2000 (triangles). The dashed blue line is BFM from
      Fig.~\ref{fig:bump}.  In (E,F), the lines correspond to BFM
      (blue) and BFM trained without the building blocks accounting
      for adverse and favorable mean pressure gradient effects
      (green).  White circles are experimental
      values\cite{bumpTN}.
  \label{fig:bumpsup}}
\end{figure}

We demonstrate that excluding BBFs modeling favorable/adverse
  pressure gradients from the training set significantly diminishes
  the performance of BFM on the Gaussian bump, as shown in
  Figures~\ref{fig:bumpsup}(E,F). The grid resolution matches that of
  Figure~\ref{fig:bump}. The green line represents a version of BFM
  with the SGS trained solely on BBFs derived from turbulent channel flows, 
  without considering BBFs that account for the effects of favorable/adverse
  pressure gradients. In the absence of these specific BBFs, the model
  fails to accurately capture flow separation. 
  These results illustrate how the BFM framework
  can systematically incorporate additional flow physics by
  integrating new BBFs, a capability not explicitly demonstrated  by previous ANN-based SGS
  models.

\subsection*{CRM-HL: details of the computational set-up}

We follow the computational setup from \cite{goc2023}.
A semi-span aircraft geometry is simulated in a hemi-sphere of radius $1000c_a$,  
where $c_a$ is the mean aerodynamic chord.
The symmetry plane is treated with free-slip and no penetration
boundary conditions.  A uniform plug flow is used at the front half of the
hemisphere.
A
non-reflecting boundary condition with specified freestream pressure
is imposed at the rear half of the hemisphere. 
Several grid refinements are considered, as illustrated in
Fig.~\ref{fig:BFMa}, being the coarsest and the smallest grid elements
$\Delta_{\max} \approx c_a$ and $\Delta_{\min} \approx 2\times 10^{-3} c_a$,
respectively.
This leads to a total number of grid points is 30 million and 
the number of grid points per boundary layer thickness ranges from zero at 
the leading edge of the wing to twenty at the trailing edge. 
The reader is referred to \cite{goc2023} for more
details about the numerical setup and gridding strategy.

%

\subsection*{Juncture flow: details of the computational set-up}

We follow the computational setup from \cite{lozanoduran2022}
The aircraft is centered in a rectangular prism whose sides are about 5 times
the fuselage length, $L$, in the three directions.
A uniform plug flow is imposed at the bottom and front boundaries;
a non-reflecting boundary condition is used at the top and rear boundaries;
and the lateral sides are modeled as free-slip.
The mesh is constructed using a Voronoi diagram with a background grid size of 
$\Delta \approx 0.04L$, and several layers of refinement around the aircraft
that leads to minimum grid size over the aircraft's surface of $\Delta_{\min}
\approx 4 \times 10^{-4}L$.
The reader is referred to ref.\cite{lozanoduran2022} for more
details about the numerical setup and gridding strategy.

\section*{Acknowledgments}
This work was supported by the National Science Foundation under grant
\#2317254 and by an Early Career Faculty grant from NASA’s Space
Technology Research Grants Program (grant \#80NSSC23K1498).  S.~C was
supported by The Boeing Company. G.~A. was partially supported by the
NNSA Predictive Science Academic Alliance Program (PSAAP; grant
DE-NA0003993).  The authors acknowledge the support of the CTR Summer
Program 2022 at Stanford University.  The authors also acknowledge the
Massachusetts Institute of Technology, SuperCloud, and Lincoln
Laboratory Supercomputing Center for providing HPC resources that have
contributed to the research results reported here.
Finally, the authors want to thank Rahul Agrawal for sharing the data
of the Gaussian bump, and G.A. wishes to thank Prof. Manuel
Garc\'{i}a-Villalba for insightful discussions.

\section*{Author contributions}
A.L.D. devised the idea and also wrote the manuscript, G.A. developed
the model and wrote the manuscript, Y.L. developed and trained
preliminary versions of the BFM, S.C. ported the BFM to GPU
architecture, K.G. run the simulations for conventional WMLES and
provided guidance on test cases and their numerical setup.

\section*{Competing interests}
The authors declare no competing interests.

\bibliographystyle{unsrt}
\bibliography{references}

\end{document}